# Avoided mode crossings in cylindrical microwave cavities

I. Stern,[1] G. Carosi,[2] N. S. Sullivan[1] and D. B. Tanner[1]
[1]Department of Physics, University of Florida, Gainesville, Florida, 32611, USA
[2]Lawrence Livermore National Laboratory, Livermore, California, 94550, USA

Abstract

Conventional axion haloscope detectors require high-$Q$ cavities with tunable TM modes whose resonant electric field occupies as much of the full volume of the cavity as possible. An analytical study of the effects of longitudinal symmetry breaking within microwave cavities was conducted to understand the mode structure. The study revealed longitudinal symmetry breaking of the cavities is the mechanism for avoided mode crossings in cylindrical microwave cavities. The results showed the size of the gaps in the search frequency spectrum due to an avoided mode crossing is roughly proportional to the magnitude of selective size of gaps between tuning mechanisms and cavity endcaps, or other deviations from perfect longitudinal uniformity of the cavity. The form factors of the modes in avoided mode crossings are equal when the frequencies of the modes are at the point of closest approach.

Introduction

The axion particle [1-3] is a theoretical cold dark matter candidate [4] that is of great interest to the science community. The most promising axion detector is known as a haloscope [5]. Mechanically tuned cylindrical microwave cavities are an integral part of convention axion haloscope detectors [6-12], though non-cavity detectors have been proposed recently [13,14]. Axion searches are conducted by measuring the power of select modes within the detector's cavity across a frequency tuning range.

Conducting or dielectric rods are used to alter the boundary conditions of cylindrical cavities. The new boundary conditions reshape the modes, shifting the mode frequencies. The tuning rods are translated inside the cavity or incrementally inserted into the cavity in the direction of the cylinder axis to obtain a large frequency range. It is important to maintain the translational geometric symmetry of the cavity [6], so the coarse tuning should be done by translating sidewise a rod which extends the full length of the cavity. Axion searches are conducted by time-integrating power measurements of a specific TM mode, shifting the mode slightly with the tuning mechanism, and repeating the process through the entire range of the microwave cavity [7].

The probability of an axion detection is dependent on the power produced inside the microwave cavity, given by [8,9]

$$P_{mnp} = 190 \text{ yW} \left(\frac{V}{136 \text{ L}}\right)\left(\frac{Q_L}{50,000}\right)\left(\frac{B_0}{6.8 \text{ T}}\right)^2 \left(\frac{C_{mnp}}{0.4}\right)\left(\frac{g_\gamma}{0.97}\right)^2 \left(\frac{\rho_a}{0.45 \text{ GeV cm}^{-3}}\right)\left(\frac{f_a}{650 \text{ MHz}}\right), \quad (1)$$

where $V$ is the volume of the cavity, $Q_L$ is the loaded quality factor of the cavity, $B_0$ is the magnitude of a static magnetic field externally applied to the cavity, and $g_\gamma$, $\rho_a$, and $f_a$ are axion parameters (see Refs. 5, 8, and 10 for addition detail of the detector). The indices $m$, $n$, and $p$ distinguish cavity modes in the typical manner. The form factor, $C_{mnp}$, is a mode specific measure of usable field within the cavity. The form factor of a mode can vary between 0 and 1, and is given by

$$C_{mnp} \equiv \frac{\left|\int d^3x\, \mathbf{E}_{mnp}(\mathbf{x}) \cdot \mathbf{B_0}\right|^2}{B_0^2 V \int d^3x\, \varepsilon(\mathbf{x}) |\mathbf{E}_{mnp}(\mathbf{x})|^2}, \quad (2)$$

where $\mathbf{E}_{mnp}$ is the electric field of the cavity search mode, and $\varepsilon$ is the relative permittivity inside the cavity.

Commonly, the magnetic field is obtained by a superconducting solenoid. In such cases, $\mathbf{B_0}$ is in the direction of the cylindrical axis of the cavity to first order. Thus, search modes with a large $E_z$ component are needed to conduct an axion search. Further, the field must be in-phase over a majority of the volume, or the volume integral in the numerator of Eq. 2 goes to zero. For this reason, generally only $TM_{mn0}$ modes yield a nonnegligible form factor. The viable search mode with the lowest frequency is the $TM_{010}$, which typically is the mode with the highest form factor.

The form factor cannot be directly measured, requiring axion haloscope experiments to rely heavily on analytical predictions. To validate the predictions, predicted mode frequencies and quality factors are compared to measurements across the tuning range of the cavity. An $N$-dimension map, known as a mode-map, displays the measured mode frequencies versus tuning mechanism location, where $N - 1$ is the number of degrees of freedom of the tuning mechanism. For example, a microwave cavity with 2 tuning rods, each with one translational motion, would yield a 3-dimensional map. The mode-map depicts mode frequencies as a function of tuning device position(s) and displays mode crossings within the tuning range of the cavity.

The mode-map is overlaid onto the predicted mode frequencies to validate the predictions and identify the search mode. Analysis is conducted to confirm that the predictions sufficiently match the measurements to validate the computed form factors. Figure 1 shows an example of a mode-map overlaid onto predicted modes. A single tuning rod is translated along an arc to change the mode frequencies; the x-axis annotates the rod location in degrees on the arc. The black lines represent analytical predictions and the red data points are frequency measurements of modes. The example shows that predictions do not always match measurements exactly.

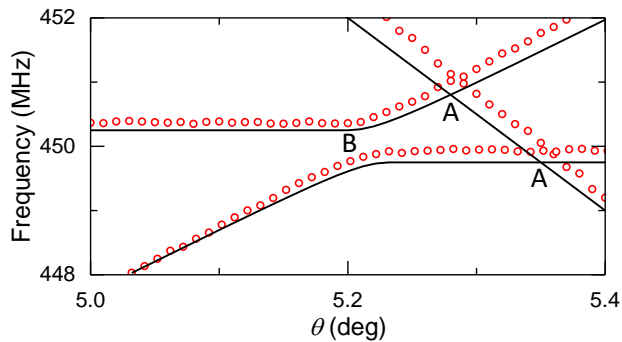

FIGURE 1: Select mode frequency measurements (red data points) and predicted frequency measurements (black lines) of a tunable microwave cavity shown across a section of the tunable range. Mode crossings are labeled "A" and the AMC is labeled "B".

Existing experiments have searched for axion particles with sufficient sensitivity in the frequency range of 460–890 MHz [10], but no discovery has been observed to date. Recently, significant effort has been made to extend the tuning range of haloscope cavities to frequencies above 1 GHz [11,12]. Such endeavors have required advancing the current technology [13,16,17,18]. In an effort to enhance the understanding of the effects from breaking longitudinal geometric symmetry within haloscope cavities, an analytical study was conducted [19]. This paper expands on findings of that study and ties the results to the two-state Hamiltonian which predicts AMC (see Ref. 19 for further findings of the study).

Coupled two-level systems

The effective Hamiltonian of an unbound, two-state system can be written as [20]

$$\mathbf{H} = \begin{bmatrix} E_1^0 - \frac{i}{2}\Gamma_1^0 & v - \frac{i}{2}\eta \\ v - \frac{i}{2}\eta & E_2^0 - \frac{i}{2}\Gamma_2^0 \end{bmatrix}, \quad (3)$$

where $E_i^0$ is the unperturbed energy level of the two states, and $\Gamma_i^0$ is the width of the unperturbed, unbound states. The off-diagonal components come from decomposition of the Hamiltonian into real and imaginary parts and represents the "amplitude" of the perturbation (see Ref. 20 for proof). The Hamiltonian could be static or condition-dependent.

In the unperturbed condition, the off-diagonal components are zero, and the energy levels of the two states can be identical. However, the Wigner and von Neumann theorem [21] shows the two energy levels cannot be identical when there is a non-vanishing off-diagonal interaction.

Equation 3 can be applied to damped oscillating modes by treating the energy state as the mode frequency ($f_i$) and the width of the unbound state as the band width of the mode,

$$\mathbf{H} = \begin{bmatrix} f_1^0\left(1 - \frac{i}{2Q_1^0}\right) & v - \frac{i}{2}\eta \\ v - \frac{i}{2}\eta & f_2^0\left(1 - \frac{i}{2Q_2^0}\right) \end{bmatrix}. \quad (4)$$

Here $Q$ is the quality factor of the damped modes. The off-diagonal components remain unchanged in the general form. Analogous to Eq. 3, the Hamiltonian would be parameter-dependent if the mode frequencies depend on a parameter (e.g., tuned oscillators).

Mode frequency tunability has been long demonstrated in optical [22], electromechanical [23,24], and microstrip [25,26] cavities. Such cavities actively adjust the frequency of one or more modes across a given band. However, cavity modes often tune at differing rates and, in some cases, some modes do not change frequency throughout the tuning range. For example, tuning techniques that significantly alter the frequencies of transverse magnetic (TM) modes in a microwave cavity typically have minimal effect on the frequencies of transverse electric (TE) and transverse electromagnetic (TEM) modes, and vice versa.

When the frequency of a tunable mode passes through the frequency of another mode, such that the modes have exactly the same frequency within the tuning range, the frequency at the intersection is a mode crossing [27]. Equation 4 suggests a mode crossing only occurs under two conditions. The first is when the perturbation amplitude is zero ($v = \eta = 0$). The second occurs when the imaginary part of the perturbation amplitude is zero ($\eta = 0$) and the real part of the perturbation amplitude meets the criterion [20]

$$|4\nu| < |Q_1^0 - Q_2^0| \frac{f^0}{Q_1^0 Q_2^0}. \tag{5}$$

Here $f^0$ is the mode crossing frequency in the unperturbed state. The equation states when the $Q$-factor of one mode is sufficiently greater than the $Q$-factor of the other, an energy exchange between the modes will not occur and the modes will not mix. The result is analogous to a lack energy exchange between two oscillators with the same resonance frequency when the damping of one oscillator is sufficiently greater than the damping of the other [28].

When the aforementioned criteria are not met, the two modes will approach the same frequency but will never be identical. Rather, a frequency gap is maintained throughout the tuning. This phenomenon is known as an avoided mode crossing (AMC) [29]. For visualization, Fig. 1 shows (A) two mode crossings and (B) one AMC. The intersections of the black lines (red data) are the two mode crossings. The region where the bends in the two black lines (red data) are closest is the AMC.

Microwave cavities have two directions of geometric symmetry: transverse and longitudinal [6]. Transverse symmetry consists of rotational invariance (discrete or continuous) or parity of the field in the transverse plane of the cavity. In cylindrical cavities, the transverse plane is normal to the cylindrical axis. Longitudinal symmetry is translational invariance between the ends of the cavity. In cylindrical cavities, the longitudinal direction is parallel to the cylindrical axis.

Breaking geometric symmetry in optical cavities has been investigated in depth [30,31,32]. Studies have discovered how symmetry breaking in Fabry-Perot cavities cause AMC [33,34,35] and demonstrated how the symmetry breaking parameters effect AMC [36,37,38]. Mode localization caused by symmetry breaking in microwave cavities has also been studied [6,12]. Mode localization is the phenomenon where a significant percentage of the stored energy resonates within a relatively small portion of the cavity.

Localization has been predicted from the eigenfunction solution for many years [39,40,41]. However, recent findings have revealed that geometric longitudinal symmetry breaking in microwave cavities has effects that go far beyond mode localization. In particular, longitudinal symmetry breaking in the cavity produces several ramifications on the cavity mode structure that are detrimental to axion dark matter haloscope searches [19].

Model for numerical simulations

A 3-dimensional finite element model (FEM) was created using COMSOL [42] version 5.0. The modeled cavity was a right circular cylinder with conducting (pure copper) boundary conditions. The model had a radius of $R = 6.826$ cm and a height of $L = 34.13$ cm. The cavity utilized a single conducting (copper) tuning rod of diameter $d = 3.631$ cm. The tuning rod is oriented parallel to the cylinder height and extends the length of the cavity in the unperturbed case (perturbations are discussed below). The rod was translated radially from the center axis to the wall. An eigenfunction solver computed the mode frequencies, quality factor, and form factor at varying locations of the tuning rod. A mesh size of approximately 1/7$^{th}$ the wavelength was used, based upon vendor-provided documentation for the FEM solver.

The simulated cavity was altered (perturbed) to address two types of longitudinal geometric symmetry breaking: mechanical gaps at the ends of the tuning rods and mechanical tilt of the tuning rod. The mechanical gap at the rod ends is common in haloscope cavity designs, as the rod must move within the cavity, and thus cannot be mechanically connected or directly electrically grounded at the endcaps. A tilt of the tuning rod is a result of mechanical misalignments. Models with mechanical gaps of $0.001L$, $0.003L$ and $0.005L$, with mechanical tilts of $0.25°$, $0.50°$, $1.00°$ and $1.80°$, and with no longitudinal symmetry breaking were evaluated. Coupled effects between the gap and tilt were not investigated, so all tilted rods were grounded flush at the endcaps. Additionally, select simulations were conducted with high-$Q$ cavities (approximately superconducting surfaces) to identify the changes of varying $Q$. Low-$Q$ cavities were not assessed.

Figure 2 shows the unperturbed cavity model used in this study in wire frame with the tuning rod (a) on center and (b) against the wall. The tuning rod on center is defined as $x = 0$, where the maximum value for $x$ is $R - d/2$. The tuning rod maintains longitudinal symmetry (i.e., no mechanical gap or tilt) in the figure. Perturbed cavity models (not shown) had either mechanical gaps located at both ends of the tuning rods, resulting in "floating" rods, or a mechanical tilt of the tuning rod in the direction out of the page in Fig. 2.

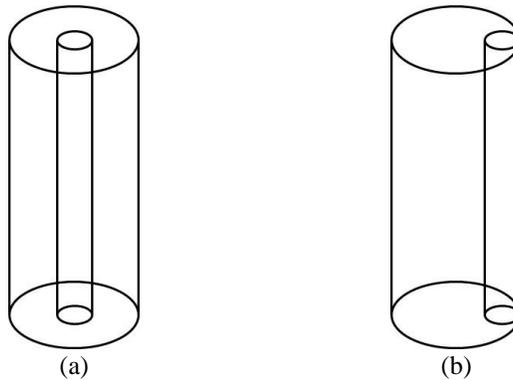

(a)          (b)

FIGURE 2: Unperturbed FE model shown in wire frame. (a) Tuning rod on center with the cavity axis. (b) Tuning rod touching the wall.

Results

Mode-maps of the different configurations exhibited AMC only when longitudinal geometric symmetry was broken. Further, the form factor of viable search modes would significantly decrease at the AMC. Figure 3 shows the mode-maps for the lowest viable search mode ($TM_{010}$) across part of the tuning range for (top panel) the cavity that maintains longitudinal symmetry, and (bottom panel) the cavity with a tuning rod end gaps of $g = 0.005L$. The displacement of the rod is shown on the x-axis, normalized by the cavity radius (analogous to the tuning parameter, $\theta$, in Fig. 1). Only a portion of the rod displacement is shown. Hatch marks indicate regions where no simulation data was obtained due to excessive computational time. The modes are labeled for clarity. The results detailed below are consistent across all simulations with mechanical gaps from the study.

The mechanical gap slightly lowers the frequency of the $TE_{214}$ mode. The $TE_{214}$ mode is a two-fold degeneracy in the top figure. When longitudinal geometric symmetry is maintained, the modes are degenerate regardless of the tuning rod location. However, breaking longitudinal symmetry lifts the degeneracy when the $TM_{010}$ mode frequency is nearly the same as the $TE_{214}$ frequency. One of the lifted degeneracy modes does not change in frequency with the translation of the tuning rod. The other lifted degeneracy forms an AMC with the $TM_{010}$ mode.

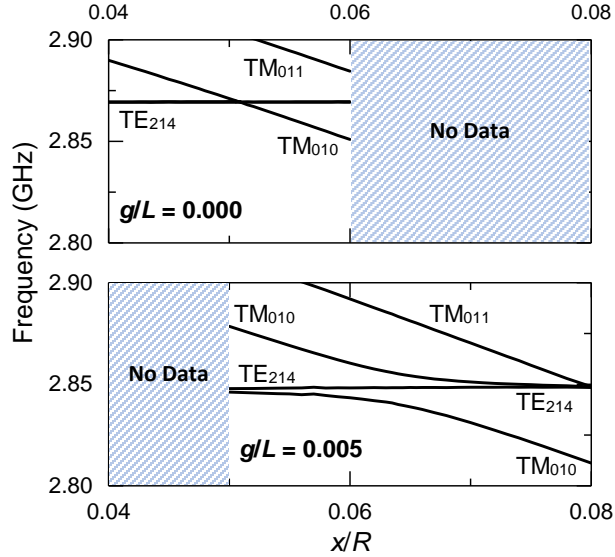

FIGURE 3: Mode crossings from the modeled cavity. (Top) Longitudinal geometric symmetry maintained. (Bottom) Mechanical gap at the rod ends of $g = 0.005L$.

The modes are pure TE/TM modes when the frequencies are sufficiently separated. But the mode that is the $TM_{010}$ at the left side of the plot converts to a degenerate $TE_{214}$ mode at the right side, while one of the degenerate $TE_{214}$ modes on the left side converts to the $TM_{010}$ mode on the right. The simulations revealed that during the AMC phase of the mode-map, the two modes form TE/TM hybrid modes.

Figure 4 shows a cross-section of the electric field of (a-b) two hybrid modes at the closest approach and (c) a coexisting TE mode with a frequency between the hybrid mode frequencies. The tuning rod is displaced from the center and is located at $x = 0.0644R$ (see Fig. 8(a)). The ends of the tuning rods have a mechanical gap of $g = 0.005L$. The arrows indicate the electric field vectors in plane; the location of the vector origin is the tail of the arrow. The color displays the magnitude of the electric field, where red indicates higher field and blue indicates lower field. The white area in the center is the location of the tuning rod.

The arrows in (a) and (b) clearly show that neither mode is a pure TE. The hybrid modes display characteristics of a cross between TM and TE modes on the left side of the cavity (as shown) but appear mostly TE-like modes on the right side. The corresponding form factor is low, with $C = 0.28$ for both modes (see Fig. 7). The maximum field shown in the cross-section is located in the tuning rod end gaps for both hybrid modes. This phenomenon is caused by a capacitive effect. Electric charge oscillates along the surface of the tuning rods and builds up at the ends in the gap, producing a voltage potential between the rod ends and the endcaps of the cavity.

Mechanical tilting of the tuning rod demonstrated similar AMC results though the effect was less pronounced. Figure 5 shows the mode-maps for the lowest viable search mode ($TM_{010}$) across part of the tuning range for (top panel) the cavity that maintains longitudinal geometric symmetry, and (bottom panel) the cavity with a mechanical tilt of $\varphi = 1.00°$. The displacement of the rod is shown on the x-axis, normalized by the cavity radius. Only a portion of the rod displacement is shown. The modes are labeled for clarity. The results detailed below are consistent across all simulations with tuning rod tilt from the study.

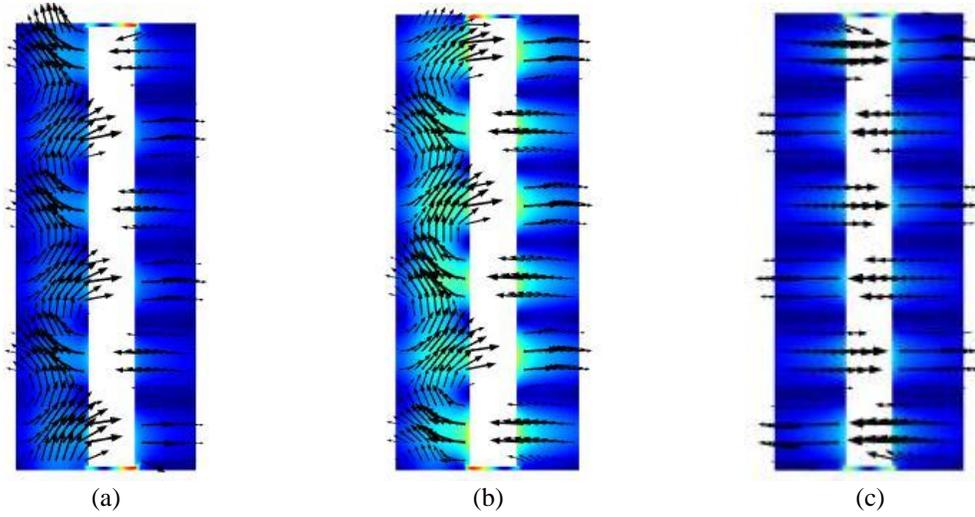

FIGURE 4: Cross-section of the electric field of (a-b) the hybrid modes and (c) the non-mixing TE mode at an AMC. The rod is displaced to the right, to $x = 0.0644R$. The field patterns correspond to the tuning rod location denoted by the vertical dashed line in Fig. 8, later in the paper.

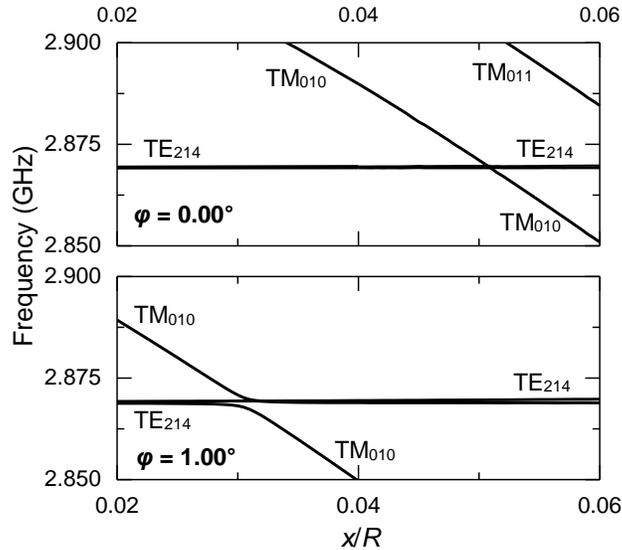

FIGURE 5: Mode crossings from the modeled cavity. (Top) Longitudinal geometric symmetry maintained. (Bottom) Mechanical tilt of the rod of $\varphi = 1.00°$.

The mechanical tilt of the tuning rod lowers the $TM_{010}$ frequency, seen as a shift left of the $TM_{010}$ mode trace. However, the tilt raises the $TM_{011}$ frequency. As a result, the $TM_{011}$ mode trace is not depicted in Fig. 5 bottom panel.

The AMC detected by the study are consistent with those predicted by Eq. 3, with longitudinal geometric symmetry breaking causing the perturbation. Equation 4 can be rewritten as

$$\mathbf{H} = \begin{bmatrix} f_1^0(\chi)\left(1 - \frac{i}{2Q_1^0(\chi)}\right) & Y(\sigma) \\ Y(\sigma) & f_2^0(\chi)\left(1 - \frac{i}{2Q_2^0(\chi)}\right) \end{bmatrix}. \tag{6}$$

$Y$ is a complex term dependent on an explicit symmetry breaking parameter, $\sigma$, where $Y(0) = 0$ when longitudinal geometric symmetry is maintained ($\sigma = 0$). The frequency and $Q$ of the modes are dependent on the tuning parameter, $\chi$. In the study, the tuning parameter is the rod location, $x/R$, but any tuning parameter is viable in Eq. 6 as long as the explicit symmetry breaking parameter constant across tuning. The explicit symmetry breaking parameter ($\sigma$) is dependent upon $g/L$ for the cavities with mechanical gaps and $\varphi$ for the cavities with mechanical misalignments.

When longitudinal geometric symmetry is maintained, no AMC exists and $f_1 = f_2$ at the mode crossing (see Fig. 5 top panel). It is significant to note the study showed transverse symmetry breaking did not result in AMC. Longitudinal geometric symmetry breaking was necessary and sufficient to cause AMC, showing longitudinal geometric symmetry breaking is the mechanism for mode mixing which produces the AMC (see below). The discovery is analogous to findings observed in optical cavities [34].

The non-mixing degenerate TE mode is unchanged by symmetry breaking in Fig. 3 and 5. This finding was consistent in the study with mode crossings occurring with degenerate modes. The complete three-state Hamiltonian can be written for the modes as

$$\mathbf{H} = \begin{bmatrix} f_1^0(\chi)\left(1 - \frac{i}{2Q_1^0(\chi)}\right) & 0 & Y(\sigma) \\ 0 & f_2\left(1 - \frac{i}{2Q_2}\right) & 0 \\ Y(\sigma) & 0 & f_3^0(\chi)\left(1 - \frac{i}{2Q_3^0(\chi)}\right) \end{bmatrix}, \tag{7}$$

where $f_2$ and $Q_2$ are constants. When $Y(\sigma) = 0$, then (1) $f_1 = f_2$ for all values of $\chi$, (2) $Q_1 = Q_2$ for all values of $\chi$, and (3) all three frequencies, $f_1$, $f_2$, and $f_3$, are equal at the mode crossing.

The simulations showed mode mixing causes AMC in microwave cavities. All modes maintained TE/TM mode purity when no symmetry breaking is present, and no AMC was observed. But, when longitudinal symmetry was broken, TE/TM hybrid modes were often observed when the frequency of TE and TM modes were approximately the same. AMC were witnessed only when hybrid modes were formed. The cause for the inconsistency in the existence of hybrid modes was not uncovered but is related to the real part of $Y$ as shown in Eq. 5. The study revealed mode hybridization and AMC were not altered by changes in $Q$ at high values, but a low-value threshold was not investigated.

The cavity form factor was significantly affected by AMC. Figure 6 shows the form factor of the lowest viable search mode (TM$_{010}$) for the modes displayed in Fig. 3: (top panel) cavity maintains longitudinal symmetry; (bottom panel) cavity with a tuning rod end gaps of $g = 0.005L$. The displacement of the rod is shown on the x-axis, normalized by the cavity radius. Hatch marks indicate regions where no simulation data was obtained due to excessive computational time. When longitudinal symmetry is maintained, the form factor is reduced slightly from transverse geometric symmetry breaking [6,11,19]. However, the form factor is greatly reduced during the AMC due to the formation of hybrid modes.

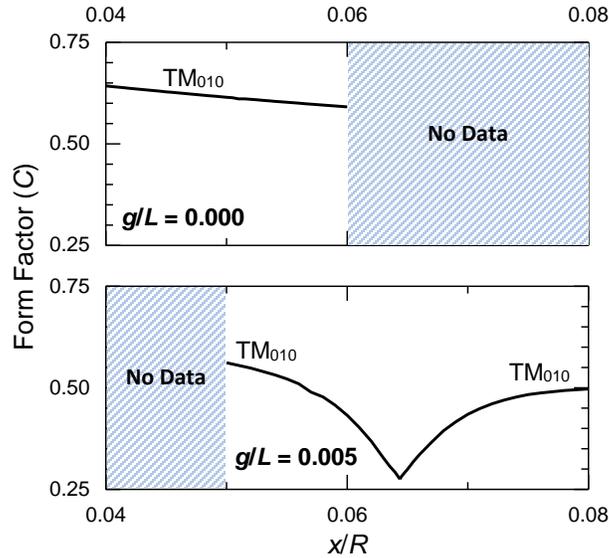

FIGURE 6: Highest form factor from the modeled cavity. (Top) Longitudinal geometric symmetry maintained. (Bottom) Mechanical gap at the rod ends of $g = 0.005L$.

As the difference between the frequencies of the modes decreases and hybrid modes are formed, the form factor of the perturbed $TM_{010}$ mode decreases as the form factor of the perturbed $TE_{214}$ mode increases. The form factors of the two modes are the same when the frequencies of the modes are closest (see Fig. 4), which was consistent with all simulations. Figure 7 shows the form factor of the two modes in Fig. 6 bottom panel across the assessed frequency tuning range. The results detailed below are consistent across all simulations from the study.

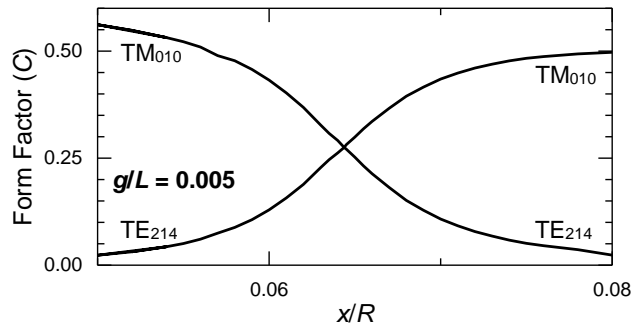

FIGURE 7: Form factor of the $TM_{010}$, $TE_{214}$, and hybrid modes in the modeled cavity with a mechanical gap at the rod ends of $g = 0.005L$.

The tracking of the axion search mode through the AMC requires switching from one hybrid mode to the other at the point when the form factors are equivalent. The shift causes a gap in the frequency spectrum of the axion search. The simulation showed that the gap in frequency spectrum due to AMC was proportional to the explicit symmetry breaking parameter.

Figure 8(a) shows the gap in the frequency spectrum ($\Delta f$) for a mechanical gap of $g = 0.005L$; the data are identical to the bottom panels of Fig. 3 and Fig. 6, but the scale is changed for clarity. Figure 8(b) shows $\Delta f$ as a function of mechanical gap. The frequency-gap is defined as the difference in frequencies between the hybrid modes at the point where their form factors are identical (see Fig. 7). $\Delta f$ is normalized by the frequency of the mode crossing ($f_{mc}$) in the unperturbed cavity (see Fig. 3 top panel), such that $f_1^0(\chi_{mc}) = f_2^0(\chi_{mc}) = f_{mc}$. The gap is normalized by the length of the cavity. The modes shown in Fig. 4 lie in line with the blue dotted line in Fig. 8(a), labeled a, b, c.

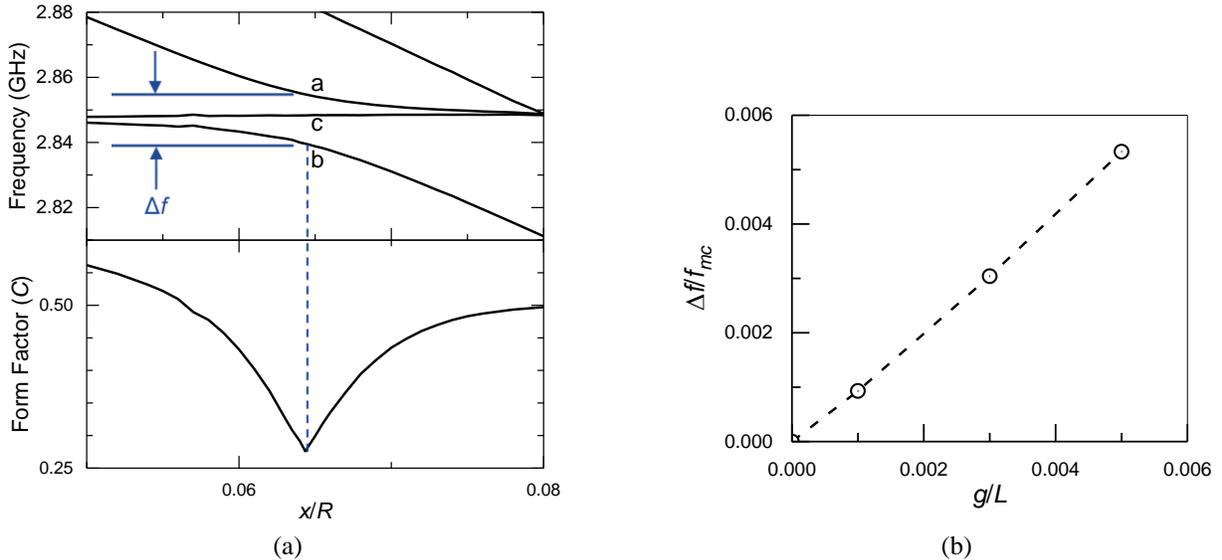

(a) (b)

FIGURE 8: Gap in frequency spectrum versus mechanical gap at rod ends. Mode structures shown in Fig. 4 corresponds to the tuning rod location denoted by the vertical dashed line and the letters correspond to the Fig. 4 labels.

The results show that, for small mechanical gaps, the relative frequency-gap size is quasi-linear with mechanical gap size. Similar results were found for mechanical tilt in the study. The results are predicted by Eq. 6, further supporting the perturbation in the two-state Hamiltonian of the modes as being a function of an explicit symmetry breaking parameter. The findings also suggest the explicit symmetry breaking parameter ($\sigma$) is greater for a mechanical gap of $g = 0.005L$ than for a mechanical tilt of $\varphi = 1.00°$ by comparing the magnitude $\Delta f/f_{mc}$ for the two cases.

Conclusion

The results of this cavity mode simulation study revealed that longitudinal geometric symmetry breaking, and not transverse geometric symmetry breaking, is the mechanism for avoided mode crossings (AMC) in cylindrical microwave cavities. Longitudinal symmetry breaking causes pure TE/TM modes to mix, forming hybrid modes at crossings. AMC are predicted only when hybrid modes are present. However, even when longitudinal symmetry breaking is present, not all TE/TM crossings resulted in hybrid modes. The factors for why some crossings mix while others do not within the same tuned cavity was not determined, but may be related to sufficiently varying quality factors of crossing modes resulting in a lack of energy exchange as suggested by Eq. 5. Gaps in the frequency spectrum of the axion search due to AMC are approximately linear with the explicit symmetry breaking parameter of the Hamiltonian on the order typical of axion haloscope detectors.

The findings strongly suggest that axion haloscope detectors minimize longitudinal geometric symmetry breaking, such as mechanical gaps at the rod ends and mechanical tilt of the rods, to increase the probability of axion detection and reduce mode complexity during searches. The ratio of the frequency spectrum gap size to the search frequency at the gap was approximately equal to the ratio of the mechanical gap size to the length of the cavity. Assuming an equal probability of an axion being observed at a given frequency within a spectrum, a spectrum gap due to an AMC of 0.5% of the frequency tuning range of a cavity would reduce the confidence limits of detection by ~0.5% for the haloscope. The limit would require a mechanical gap of

$$g \lesssim 0.005 \frac{L}{\gamma} \frac{df}{f_{mid}} , \qquad (8)$$

where $\gamma$ is the number of tuning rods in the cavity, $df$ is the frequency tuning range of the cavity, and $f_{mid}$ is the mid-frequency value of the tuning range.

Acknowledgements

The first author acknowledges support by the U.S. Department of Defense through the National Defense Science and Engineering Graduate Fellowship Program and the National Aeronautics and Space Administration through the Florida Space Research Program. This work was supported in part by the Department of Energy under Grant No. DE-SC0010280 and Grant No. DE-SC0010296 at the University of Florida.